# Approaching physical limits of latent dimensionality in optical computing

Zhenyu Zhao[1], Zijun Qiu[1], Xuan Hu[1], Yao Zhou[1], Jinlong Xiang[1], Youlve Chen[1], Chaojun Xu[1], Yuchen Yin[1], Tao Lin[2], Yikai Su[1], and Xuhan Guo[1,*]

[1]State Key Laboratory of Photonics and Communications, School of Information Science and Electronic Engineering, Shanghai Jiao Tong University; Shanghai, 200240, China.

[2]North Ocean Photonics Co. Ltd.，Shanghai, 201306, China.

*Corresponding author. Email: guoxuhan@sjtu.edu.cn.

**Abstract**

The physical implementation of artificial intelligence requires mapping computational processes onto the dynamic physical processes of the underlying computing platform. The photonic processors offer an intrinsically parallel and low-energy framework for this mapping, however, a mismatch between the potential computing capability of a bounded optical domain and the human-accessible manipulation range sets a hard integration-density ceiling on existing architectures. Here, we address this challenge by investigating the integration density limits in photonic processors through exploring the fundamental physical limits on the latent dimensionality for maximum expressivity of a bounded optical domain. These physical limits potentially serve as universal metrics for evaluating optical computing capacity. To validate these, we design and realize ultracompact multimode photonic processors approaching these limits: a 2.2 μm × 8 μm processor achieves 86.7% accuracy in experiment for iris flower classification, and a 20.6 μm × 44.8 μm processor reaches 92.9% accuracy in handwritten digit recognition. Finally, we scale this architecture to highly complex tasks by implementing a generative diffusion model for image synthesis. By grounding photonic processor design in the wave-physics origin of latent dimensionality, our results supply the missing theoretical reference point for optical computing architecture.

## Introduction

The rapid advancement of artificial intelligence (AI) is revolutionizing industries[1], yet its growth faces a critical bottleneck: escalating energy demands. While the substantial power consumption of training large-scale models constrains data centers[2], the limited energy density of batteries further restricts edge AI applications[3]. To mitigate these energy inefficiencies among these, photonic processors which leverage electromagnetic waves as low-loss information carriers, have emerged as a highly promising solution to conventional transistor-based digital computing[4]. Unlike electronic circuits plagued by resistive losses, electromagnetic waves exhibit significantly lower attenuation, a principle already validated by optical communication systems[5]. However, unlike in communication systems, using light for computing remains at early stage due to limited control over electromagnetic wave and their strong dependence on electronic circuits, which hides the full benefits of photonic processors. Despite advanced fabrication techniques, photonic processors with matrix dimensions comparable to electrical analog computing circuits typically occupy ~100× greater area[6,7].

This scaling challenge primarily stems from a potential fundamental limit of integration density. In Mach–Zehnder interferometer (MZI) mesh[8,9,6,10] architectures, each tunable element requires a footprint of tens of micrometers due to low manipulation efficiency[11], so a matrix comparable to those used in electronic analog circuits occupies roughly two orders of magnitude more chip area[6,7]. Recent inverse-designed and structurally patterned processors have improved this substantially[12–14], demonstrating compact nanophotonic processors with high computational density. Yet none of these advances has established a physical limit on how dense and scalable a photonic processor can be. Without such a bound, designers cannot know whether any given architecture is operating near the physical ceiling or whether orders-of-magnitude improvement remains available.

In this work, we resolve this scalability barrier by investigating a fundamental relationship between the latent dimensionality for maximum expressivity of a bounded optical domain and integration density limit in optical computing. We demonstrate the multimode photonic processors that achieve near-physical-limit integration density. By compressing the latent dimensionality to approach the information dimensionality being processed, we design multimode processors that reach this theoretical bound without compromising performance: a 2.2 μm × 8 μm processor on SOI platform performs three-category Iris dataset classification, while a 20.6 μm × 44.8 μm processor with 65 modes handles 64-pixel handwritten digit recognition. Furthermore, we implement multimode photonic processors in a generative diffusion model, demonstrating the capability for complex machine learning tasks. These advances prove the feasibility of optical computing devices approaching the physical density limit, paving the way for exploring space-efficient AI hardware.

## Results

### Physical Limit on Latent Dimensionality

The term "latent dimensionality" refers to the maximum theoretically achievable information processing dimensionality within a bounded optical domain, independent of specific structural implementations. Prior studies have established photonic devices' capacity for universal matrix-vector multiplication[12,8] which is a fundamental operation underpinning modern computing, particularly for machine learning applications[15]. However, compared with the traditional transistor-based analog computing systems[7], integration density remains a fundamental bottleneck for optical processors despite their potential for higher energy efficiency[16]. Consequently, investigating the physical limits of integration density for optical processors is critical for advancing optical computing systems research.

According to information theory[17], information loss occurs when the number of processing channels is fewer than the dimensionality of the input information. To ensure the functionality of optical processors, the latent dimensionality must be greater than or equal to the information dimensionality. This establishes the latent dimensionality as a fundamental physical limit for optical processing systems.

In integrated photonic platforms, the number of information-processing dimensions at a single wavelength is determined by the allowable wave numbers on the cross-section of photonic devices[18]. Conventional calculations assume pre-defined structures, such as fixed waveguides, where the latent dimensionality equals the number of supported orthogonal modes[18]. Here, we extend this framework to arbitrary cross-sectional geometries. Using effective medium theory, we show that the latent dimensionality for such structures corresponds to the number of modes supported by a homogeneous material with the highest refractive index occupying the same cross-section (Supplementary Note 1). Analogous to established mode-counting methods for slab waveguides and optical fibers[18], we derive the latent dimensionality ($Q$):

$$Q \approx \left\lfloor \frac{2W}{\lambda} \sqrt{n_{\text{high}}^2 - n_{\text{low}}^2} \right\rfloor \tag{1}$$

where $W$ is the width of the cross-section, $\lambda$ is the vacuum wavelength, $n_{\text{high}}$ and $n_{\text{low}}$ define the upper and lower bounds of the effective refractive index in the cross-section. We restrict our analysis to a single wavelength and polarization, as field modulation depends solely on refractive index modulation in this regime[11]. For multi-wavelength or multi-polarization cases, refractive index variations are typically strongly coupled[19], leading to severe crosstalk between information channels operating at different wavelengths and polarizations when performing information processing, unlike the simpler case of passive transmission. Equation (1) reveals that the latent dimensionality $Q$ scales positively with the cross-sectional width $W$. Moreover, information theory imposes a fundamental lower limit on $W$ based on the dimensionality of the input information to be processed, as previously discussed.

Figure **1a** analyzes the information dimensionality $N$ and latent dimensionality $Q$ of existing optical processing devices. For the diffractive blocks[20,21], the diffractive process first maps input modes to the full set of slab waveguide modes (defining the latent dimensionality), then compresses back to output modes. However, these blocks typically require large widths to satisfy diffraction theory, resulting in a latent dimensionality far exceeding the information dimensionality. For the MZI mesh[8–10,6], the device cross-section consists of discrete waveguides, where the number of supported modes matches the information dimensionality. Yet, the latent dimensionality, which is determined by the full spatial extent of the mesh, remains significantly larger. For ring resonators[22,23], wavelength multiplexing is typically

employed, with distinct rings designed for different wavelengths. Here, the latent dimensionality across the entire cross-section still vastly exceeds the information dimensionality, similar to MZI meshes. For the scattering blocks[12–14,24,25], which should have higher integration density due to their high efficiency of modulating light[13], the latent dimensionality is still larger than the dimensionality of input and output because of the limitation of input and output ports which require the decoupling for different waveguides (Supplementary Note 2). To deal with that, we introduce multimode processors in this work and make the latent dimensionality close to the information dimensionality, thus approaching the integration density limit. As illustrated in Fig. **1b**, if the latent dimensionality falls below the information dimensionality, information loss may occur, rendering the photonic processors non-functional. This also underscores the existence of fundamental physical limits.

Based on the preceding analysis and guided by recent breakthroughs in multimode photonic processors[26], we establish a fundamental integration density limit for unitary photonic processors (Supplementary Note 3):

$$\text{Integration Density Limit} = \frac{Q}{\text{Area}} \approx \frac{2\Delta n\sqrt{n_{high}^2 - n_{low}^2}}{\sqrt{N}\lambda^2} \quad (2)$$

where $\Delta n$ represents the effective refractive index perturbation for light modulation in the system. The unit of this limit is $\text{Dimensionality}\cdot\text{m}^{-2}$, where $\text{Dimensionality}$ corresponds to the size $N$ of the unitary matrix in the vector-matrix multiplication process. This limit provides a potential metric for evaluating high computational density photonic processors. Accordingly, we define the integration density for photonic processors as $\frac{N}{\text{Area}}$. Equation (2) reveals that improving integration density requires either high-contrast integrated photonic platforms or shorter laser wavelengths. However, both approaches demand more stringent control over structural features. An alternative strategy involves increasing $\Delta n$, which constitutes the key advantage of diffractive blocks[20] and scattering blocks[12–14,24,25] with passive structures. For devices designed to approach the integration density limit, the latent dimensionality $Q$ (which approaches the information dimensionality being processed) determines the minimum cross-section $W$, thereby establishing the maximum achievable integration density.

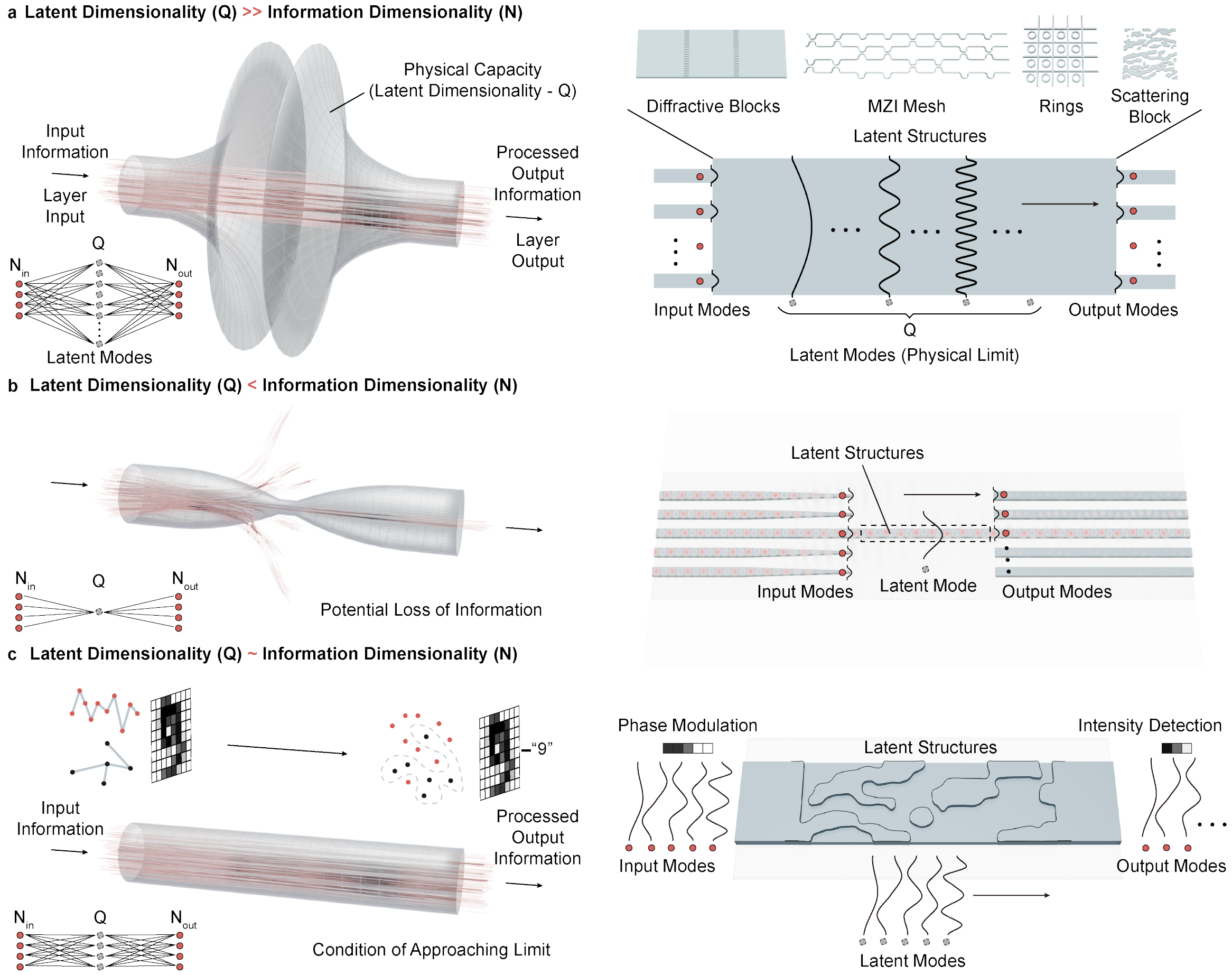

**Fig. 1 | Latent dimensionality in photonic processors. a** Schematic and analysis of conventional photonic processors, where the latent dimensionality ($Q$) typically far exceeds the information dimensionality ($N$). **b** Schematic for the condition that $Q$ is smaller than $N$, where loss of information potentially happens. **c** Schematic and analysis of our multimode photonic processor, where $Q$ approaches $N$, thus approaching the fundamental integration density limit.

## Multimode Processors Approaching the Physical Limit on Latent Dimensionality

The multimode processors approaching the physical limit on latent dimensionality is realized through inverse-designed nanophotonic media implemented on a single multimode waveguide, where previous photonic processors operated exclusively with low-order discrete modes, we achieve full cross-section mode utilization. The shallow-etched perturbation structures maintain support for all latent dimensions while modifying mode coupling characteristics, preserving the full dimensionality capacity of the unperturbed waveguide. As shown in Fig. **1c**, information encoding and processing utilize a complete set of modes matching the system's latent dimensionality, with computational operations implemented through controlled inter-mode coupling. For experimental simplification, as shown in Fig. **2a**, input vectors (representing dataset features) are encoded in the phase profiles of these modes, while output information is decoded from their intensity distributions. In the Iris flower classification task, class prediction is determined by identifying the mode order exhibiting maximum intensity, where each mode index corresponds to a specific output class. This approach demonstrates efficient dimensionality utilization while maintaining processing capability.

The laser source in this work operates at a wavelength of 1550 nm, resulting in a processor width $W$ of 2.2 μm for supporting five modes under TE polarization. To achieve perturbation structures with high degree of freedom for design, nanoscale fabrication is essential. Additionally, fabrication resolution must be carefully considered to ensure accurate alignment between the fabricated and designed structures. To address these challenges, we employ a fabrication-constrained inverse design method[27] achieving a minimum feature size of 130 nm for the designed structures (Supplementary Note 4).The design process consists of two stages: a continuous stage, where the refractive index within the design region varies continuously; a discrete stage, where the refractive index is constrained to real material values, and the material boundaries are optimized to enhance performance while adhering to fabrication constraints. A parameter $k$ is introduced during the continuous stage to progressively increase the discretization level of the refractive indices (Supplementary Note 4). Fig. **2b** displays the structural evolution during the training process at iterations 9, 49, and 95. Following the continuous stage, a conversion step bridges the two stages with the resulting level set function for the discrete stage illustrated in Fig. **2c**. Figure **2d** tracks the loss function for the iris flower classification task and the fabrication constraint loss during training, while Fig. **2e** presents the corresponding evolution of classification accuracy.

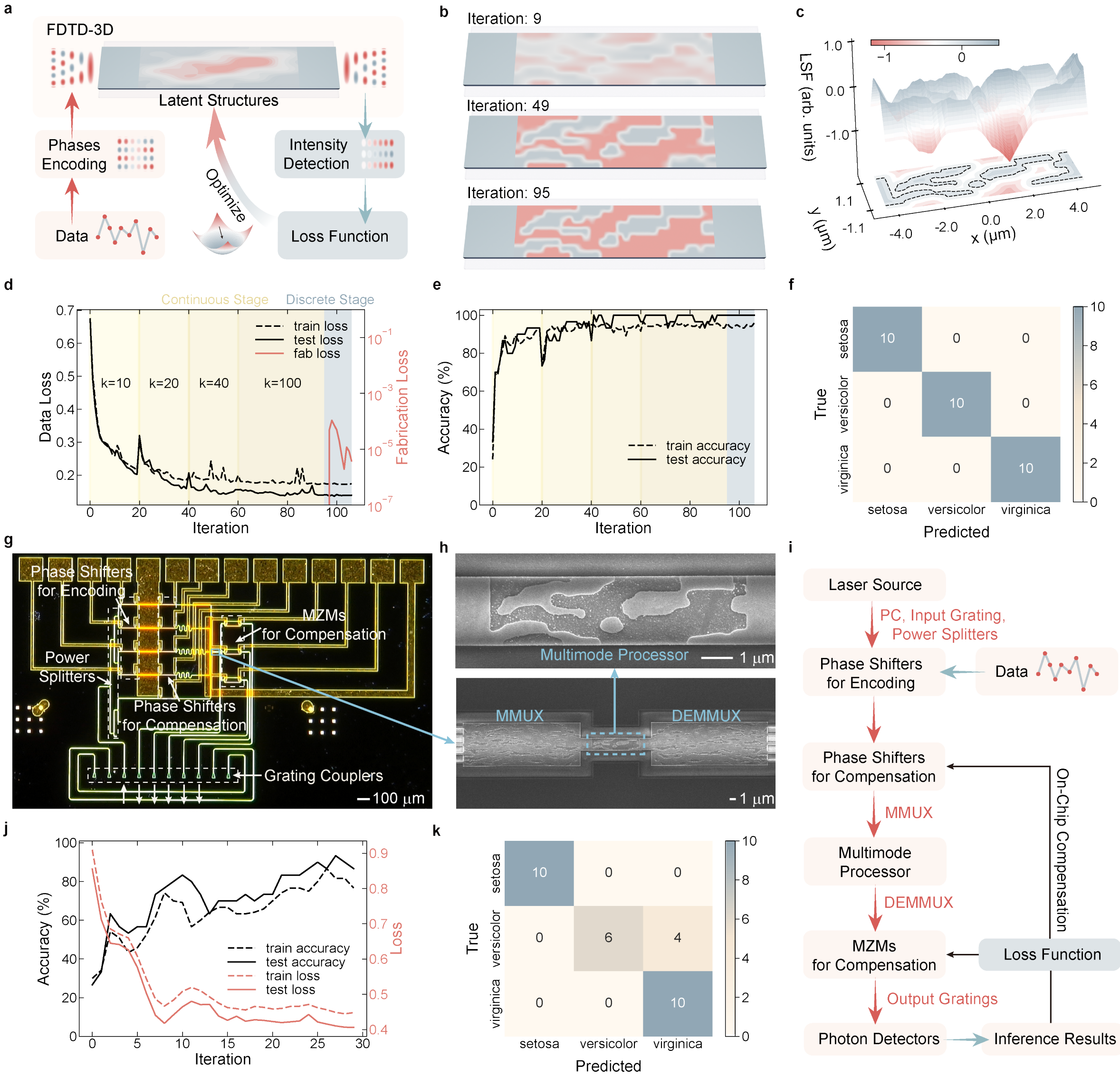

**Fig. 2 | Design, simulation and experimental results of a multimode photonic processor for slightly nonlinear iris flower classification. a** Design workflow of the multimode processor: input feature data from the training set modulate the phases of different-order modes, while output information is encoded in the mode intensities. **b** Evolution of the designed potential structures at iterations 9, 49, and 95. **c** Mapping from continuous potential structures to discrete structures after iteration 95. **d** Training progression of the loss functions: Data Loss (iris classification task) and Fabrication Loss (fabrication constraints). Iterations 0–95 correspond to the continuous stage; 96–106 to the discrete stage. **e** Evolution of classification accuracy during training. **f** Confusion matrix for the test dataset post-training. **g** Microscope image of the

fabricated photonic system. **h** SEM images of the fabricated multimode photonic processor, mode multiplexer (MMUX), and mode demultiplexer (DEMMUX). **i** On-chip compensation process flow chart. **j** Accuracy and loss function evolution during on-chip compensation. **k** Post-compensation confusion matrix for the test dataset. (PC: polarization controller; MZMs: Mach-Zehnder modulators.)

The iris flower classification dataset[28] comprises four input features per sample. We utilize five modes to introduce a slightly nonlinear processing, which is recently researched for enabling nonlinear processing in linear systems[29–32]. This nonlinear enhancement enables our model to achieve 100% test accuracy post-training, as demonstrated in Fig. **2e** and Fig. **2f**. Additionally, to account for imperfections in the designed mode multiplexer and demultiplexer (Supplementary Note 5), we incorporate their transfer functions during the training process as described in the Methods.

Figure. **2g** displays a microscope image of the fabricated chip housing the multimode photonic processor, whose scanning electron microscopy (SEM) image is shown in Fig .**2h**. In addition to the phase shifters used for encoding input information onto the mode phases, we integrate supplementary phase shifters to enable on-chip compensation for phase noise induced by waveguide imperfections. Similarly, additional Mach-Zehnder modulators (MZMs) are incorporated before the output grating couplers to compensate for imbalances among them. After compensation, the voltages applied to these components are fixed, allowing the chip to directly process input data and generate results encoded in the output light intensity, which is a fully analog computation without digital intervention. The inference and compensation workflow is illustrated in Fig. **2i**. The loss function, defined as the normalized mean squared error (NMSE) as described in Methods, is computed from the intensities detected by photon detectors. To optimize the compensation voltages for the phase shifters and MZMs, we derive gradient information from the training dataset using a finite-difference method within the experimental setup, which could also be categorized as an approach that aligns with in-situ training methodologies[21]. The Adam algorithm[33] is employed for compensation, achieving a test dataset accuracy of 86.7% after 30 iterations. The evolution of accuracy and loss on both training dataset

and test dataset are shown in Fig. **2j** and the confusion matrix on test dataset after the compensation process is shown in Fig. **2k**. While this experimental accuracy is lower than the 100% simulated performance, the discrepancy likely stems from uncorrected imperfections in the mode multiplexer and demultiplexer (Supplementary Note 5). Nevertheless, the achieved accuracy matches that of prior passive-structure implementations[13]. The integration density of this design is calculated as 0.28 $\mu m^{-2}$, which is close to the theoretical limit is 0.29 $\mu m^{-2}$ derived from Equation (2) (Supplementary Note 3).

To further validate the scalability of the multimode photonic processor and probe the integration density limit, we designed a larger-scale device with a 20.6 μm×44.8 μm footprint tailored for handwritten digit recognition. This device supports an input dimensionality of $N = 65$, achieving an integration density of 0.07 $\mu m^{-2}$, closely approaching theoretical limit of 0.08 $\mu m^{-2}$ (Supplementary Note 3). A schematic of the design is depicted in Fig. **3a**. Similar to the iris classification task, input information is encoded in the phases of input modes, while output results are extracted from the intensities at the output ports. The dataset[34] consists of 3,823 training images and 1,797 test images, each comprising 64 pixels. A 65$^{th}$ mode is introduced to enable slightly nonlinear mapping. The training process follows a two-stage approach: continuous stage and discrete stage, mirroring the methodology used for iris classification. A 130 nm minimum feature size constraint is also adopted in the training process (Supplementary Note 4).

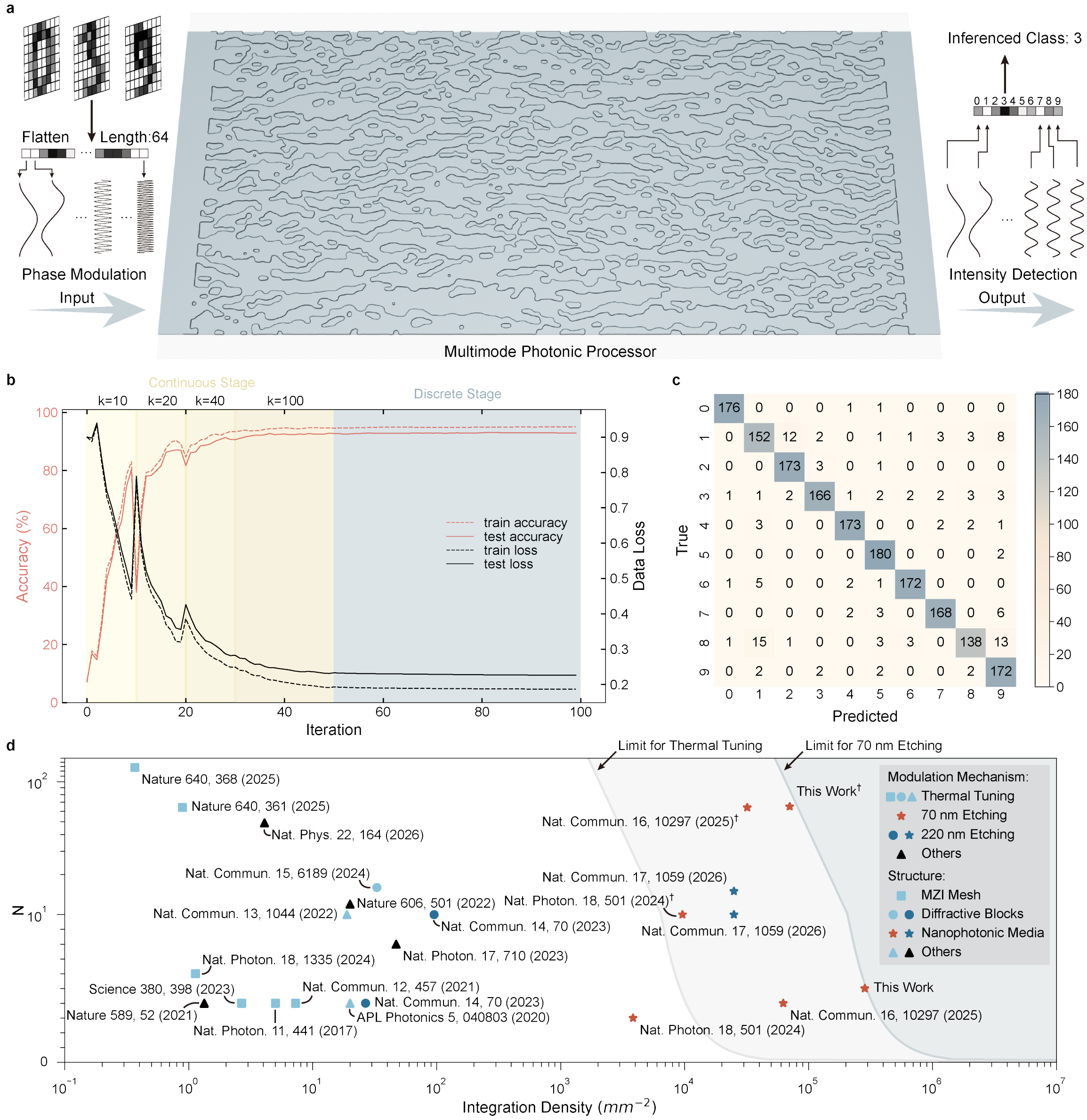

**Fig. 3 | Handwritten digit recognition using a multimode photonic processor and integration density analysis. a** Schematic of the processor architecture for handwritten digit recognition. **b** Confusion matrix showing classification performance on the test dataset after training. **c** Training evolution of prediction accuracy and loss function, with the continuous optimization stage (iterations 0-49) and discrete optimization stage (iterations 50-99) indicated. **d** Comparison of information dimensionality and integration density for integrated optical computing devices[6,9,10,12–14,20,21,26,35–42], thus annotating integration density limits for 220 nm top silicon on SOI with thermal tuning and 70 nm etching at a wavelength of 1550 nm. The y-axis uses a linear scale for N = 0-10 and a logarithmic scale for N > 10. †Simulation results.

After 100 training iterations, the system achieves 92.9% inference accuracy on the test dataset, with the corresponding confusion matrix presented in Fig. **3b**. The evolution of both loss and accuracy during training is shown in Fig. **3c**. Compared to previous implementations[13] using the same dataset, our design demonstrates both higher integration density and marginally improved accuracy. This performance enhancement may be attributed to the increased number of training iterations and the greater nonlinearity of our mapping function.

Figure **3d** compares the integrated optical computing devices in terms of information dimensionality and integration density (the raw data and calculations are provided in Supplementary Note 6). The integration-density limits for optical computing on a 220-nm-thick top-silicon SOI platform at λ = 1550 nm are plotted according to Equation (2), considering both the thermal-tuning mechanism and a 70 nm etching. As shown in Fig. **3d**, the higher modulation efficiency achieved by the etched patterns[12–14] enables the devices to surpass the integration-density ceiling imposed by thermal tuning, thereby improving scalability. The limit-approaching performance in this work originates from deliberately targeting the physical limits on the latent dimensionality through multimode manipulation. Additionally, there is a gap between existing thermal tuning devices and the physical limit. Some thermal tuning strategies with diffractive blocks[21,39] have suppressed most MZI meshes regarding the integration density. These indicates that structural optimization would be available to greatly improve the integration density with thermal tuning such as refined heaters distribution and structure design. Moreover, the free region observed between the nanophotonic media structures and the other structures indicates that feasible designs can balance integration density against other metrics. Notably, the fundamental distinction between the thermal tuning limit and the 70 nm etching limit lies in modulation efficiency rather than tunability. Incorporating phase change materials into the nanophotonic media structures[43] would simultaneously preserve high integration density and functional tunability.

**Generative Diffusion Model with Multimode Photonic Processors**

Generative models represent a transformative paradigm reshaping multiple domains, spanning images[44], audio[45], video[46], text[47] and even AI agents[48]. These models have demonstrated unprecedented capabilities compared to conventional approaches, but their increased complexity intensifies computational demands[49]. While recent work has explored optical solutions for generative models[50,51], yet the integration density is not sufficiently discussed. Here, we leverage our multimode photonic processor architecture to demonstrate the potential of photonic computing in realizing generative models.

Figure **4a** illustrates the inference and training procedure of our photonic denoising diffusion probabilistic model (DDPM)[44] for handwritten digit generation. The key difference from the conventional DDPM models with U-Net[52] is the utilization of multimode photonic convolution layers (Supplementary Note 7), which can be implemented with the structures discussed above. In these layers, the input information is encoded in the phase of modes of computational kernels, while output information is extracted from the intensity of output modes. Additional details regarding the training methodology are provided in the Methods.

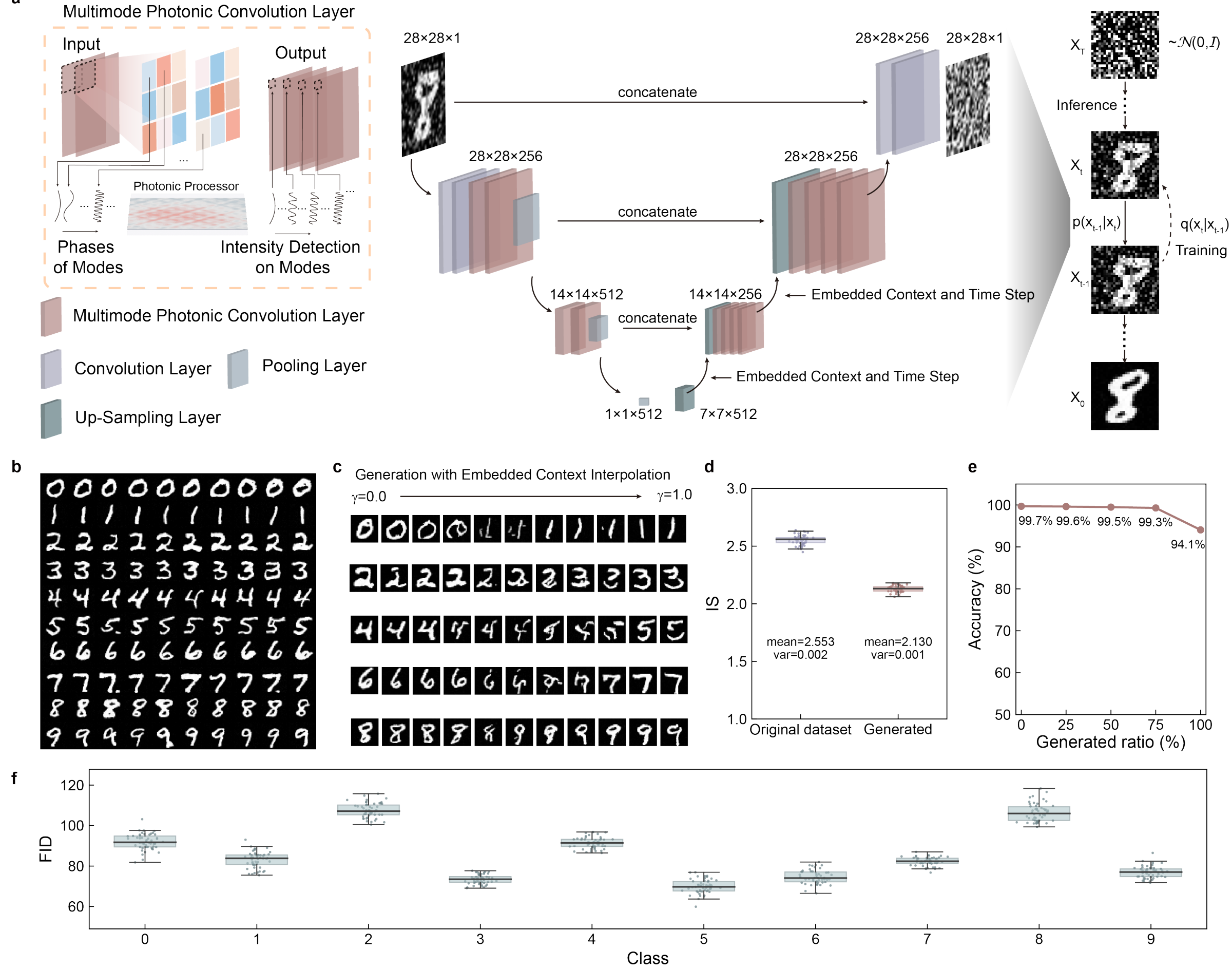

**Fig. 4 | Generative diffusion model implementation using multimode photonic processors.** **a** Architecture schematic highlighting the multimode photonic convolution layers within the diffusion model framework. **b** Representative handwritten digit images generated by the trained model. **c** Context-interpolated digit generation demonstrating continuous transitions between classes. **d** Inception Score (IS) comparison between generated images and the original MNIST dataset. **e** Classification accuracy on test data when augmenting training sets with varying proportions of generated images (0-100%). **f** Fréchet Inception Distance (FID) evaluation per digit class.

After training on the MNIST dataset[53], we employed the model in inference mode to generate digit images as shown in Fig. **4b**. Interpolated samples demonstrating continuous transitions between digit classes is also displayed in Fig. **4c**. To quantitatively assess generation quality, we computed inception score (IS)[54] and the Fréchet inception distance (FID)[55] on 50,000 generated images with a batch size of 1000, with random Gaussian noise sampling controlled by fixed random seeds. The generated images achieved a mean IS slightly below that of the original dataset as shown in Fig. **4d**. This might be due to both limited training iterations and the photonic network's currently constrained expressive capacity relative to highly optimized digital neural networks. For the FID evaluation, we show the performance on each class of generated images in Fig. **4f**. The FID values are mostly below 100, demonstrating generation quality within the acceptable range. To assess the practical utility of generated images, we trained five classifiers using datasets containing varying ratios of synthetic to original training images (0-100%). These classification models are blindly evaluated on the standard MNIST test dataset with the classification accuracies presented in Fig. **4e**. When using a dataset consisting of 75% native data and 25% generated data, the test accuracy drops by only 0.4% compared to using the fully native dataset, demonstrating the effectiveness of using generated data as a supplement to the original dataset. However, when using a fully generated dataset, the test accuracy reaches 94.1%, which, while acceptable, is likely attributable to domain shift[56]. It is expected that this issue can be further mitigated as the quality of generated images continues to improve through ongoing engineering enhancements to the model.

The multimode photonic convolution layers can be implemented using the multimode photonic processors, enabling an estimated minimum design area of approximately 2.2 $\mathrm{mm}^2$ for these layers in the generative diffusion model based on the integration density limit. This represents an exceptionally compact footprint for computational devices. Nevertheless, this estimate remains approximate because it doesn't fully account for the non-ideal unitary constraints in the photonic layers and omits scaling considerations for peripheral components (Supplementary Note 3).

## Discussion

Photonic processor design has proceeded without a principled answer to the question of how much computation a bounded optical domain can support until now.

This work introduces and demonstrates the fundamental physical limits on latent dimensionality for maximum expressivity of a bounded optical domain with inverse designed multimode photonic processors. A potential rule governing the integration density limit is summarized, which may serve as a useful benchmark for evaluating optical computing devices. For different task scales, this integration density limit can be calculated based on the required dimensionality of the information dimensionality. However, approaching this limit is not always advantageous in practical implementations, as higher-order modes exhibit increased mode dispersion, complicating system design and noise control due to additional device requirements. Novel devices capable of direct high-order mode manipulation[57] could help mitigate these challenges.

To surpass this integration density limit, wavelength and polarization modulation techniques may offer viable solutions. Nevertheless, the underlying refractive index perturbation mechanism in our theoretical framework suggests that fully decoupling the effects of refractive index variations across different wavelengths and polarizations remains challenging. Independent modulation of the nonlinear coefficients[58] could serve as a potential solution by further leveraging the wavelength domain for computing. Moreover, resonance might enhance the modulation effects of $\Delta n$ that are not captured by the theoretical framework. By exploiting resonance in the scattering blocks, one could surpass the integration-density limit, albeit at the cost of functional bandwidth.

The nonlinearity in this work arises from interference-based effects and intensity detection mechanisms, which have recently been explored as a means to achieve nonlinear computation within linear photonic circuits. Prior demonstrations of higher-order nonlinearities using this approach[29–31] suggest its potential for improving the accuracy of photonic processors in handling complex tasks

As the proposed multimode photonic processors are fundamentally analog in nature, some operations in the modern digital computers might be difficult to be realized directly, for example, the max-pooling operation[59] in the convolution neural networks. Therefore, we try to incorporate the phase-input intensity-output and complex-processing computation block, which could be realized using relatively simple photonic components while maintaining compatibility with digital architectures. However, the phase-input and intensity-output and complex-processing computation block imposes some constraints on processing capability compared to conventional neural network architectures, highlighting the need for further research on optimized hybrid implementations.

In conclusion, while integrated photonic processors hold significant potential for low-power machine learning, their ultimate performance will be governed by fundamental physical limits. Therefore, a deep understanding of the constraints on latent dimensionality and integration density is essential to overcome scalability barriers and unlock the full potential of high-performance photonic AI systems. We hope this work provides a primary pathway toward future photonic AI systems by demonstrating a near-physical-limit architecture.

## References

1. Piccialli, F. *et al.* AgentAI: A comprehensive survey on autonomous agents in distributed AI for industry 4.0. *Expert Systems with Applications* **291**, 128404 (2025).
2. Sheng, Y. *et al.* Power for AI Data Centers: Energy Demand, Grid Impacts, Challenges and Perspectives. *Energies* **19**, 722 (2026).
3. Mao, Y., Yu, X., Huang, K., Angela Zhang, Y.-J. & Zhang, J. Green Edge AI: A Contemporary Survey. *Proc. IEEE* **112**, 880–911 (2024).
4. Fu, T. *et al.* Optical neural networks: progress and challenges. *Light Sci Appl* **13**, 263 (2024).
5. Miya T., Terunuma Y., Hosaka T., & Miyashita T. Ultimate low-loss single-mode fibre at 1.55 µm. *Electronics Letters* **15**, 106–108 (1979).
6. Ahmed, S. R. *et al.* Universal photonic artificial intelligence acceleration. *Nature* **640**, 368–374 (2025).

7. Ambrogio, S. *et al.* An analog-AI chip for energy-efficient speech recognition and transcription. *Nature* **620**, 768–775 (2023).

8. Clements, W. R., Humphreys, P. C., Metcalf, B. J., Kolthammer, W. S. & Walsmley, I. A. Optimal design for universal multiport interferometers. *Optica* **3**, 1460 (2016).

9. Shen, Y. *et al.* Deep learning with coherent nanophotonic circuits. *Nature Photonics* **11**, 441–446 (2017).

10. Hua, S. *et al.* An integrated large-scale photonic accelerator with ultralow latency. *Nature* **640**, 361–367 (2025).

11. Sun, H., Qiao, Q., Guan, Q. & Zhou, G. Silicon Photonic Phase Shifters and Their Applications: A Review. *Micromachines* **13**, 1509 (2022).

12. Nikkhah, V. *et al.* Inverse-designed low-index-contrast structures on a silicon photonics platform for vector–matrix multiplication. *Nat. Photon.* **18**, 501–508 (2024).

13. Zhao, Z. *et al.* High computational density nanophotonic media for machine learning inference. *Nat Commun* **16**, 10297 (2025).

14. Sved, J. *et al.* Inverse-designed nanophotonic neural network accelerators for ultra-compact optical computing. *Nat Commun* **17**, 1059 (2026).

15. Jordan, M. I. & Mitchell, T. M. Machine learning: Trends, perspectives, and prospects. *Science* **349**, 255–260 (2015).

16. Shekhar, S. *et al.* Roadmapping the next generation of silicon photonics. *Nat Commun* **15**, 751 (2024).

17. C. E. Shannon. A mathematical theory of communication. *The Bell System Technical Journal* **27**, 379–423 (1948).

18. Allan W. Snyder, John D. Love. *Optical Waveguide Theory*. (1983). doi:10.1007/978-1-4613-2813-1.

19. Pendry, J. B., Schurig, D. & Smith, D. R. Controlling Electromagnetic Fields. *Science* **312**, 1780–1782

(2006).

20. Fu, T. *et al.* Photonic machine learning with on-chip diffractive optics. *Nat Commun* **14**, 70 (2023).
21. Cheng, J. *et al.* Multimodal deep learning using on-chip diffractive optics with in situ training capability. *Nat Commun* **15**, 6189 (2024).
22. Tait, A. N. *et al.* Microring Weight Banks. *IEEE J. Select. Topics Quantum Electron.* **22**, 312–325 (2016).
23. Huang, C. *et al.* A silicon photonic–electronic neural network for fibre nonlinearity compensation. *Nat Electron* **4**, 837–844 (2021).
24. Khoram, E. *et al.* Nanophotonic media for artificial neural inference. *Photon. Res.* **7**, 823 (2019).
25. Qu, Y. *et al.* Inverse design of an integrated-nanophotonics optical neural network. *Science Bulletin* **65**, 1177–1183 (2020).
26. Onodera, T. *et al.* Arbitrary control over multimode wave propagation for machine learning. *Nat. Phys.* **22**, 164–171 (2026).
27. Vercruysse, D., Sapra, N. V., Su, L., Trivedi, R. & Vučković, J. Analytical level set fabrication constraints for inverse design. *Scientific Reports* **9**, (2019).
28. Fisher,R. A. Iris. UCI Machine Learning Repository https://doi.org/https://doi.org/10.24432/C56C76 (1988).
29. Wanjura, C. C. & Marquardt, F. Fully nonlinear neuromorphic computing with linear wave scattering. *Nat. Phys.* **20**, 1434–1440 (2024).
30. Xia, F. *et al.* Nonlinear optical encoding enabled by recurrent linear scattering. *Nat. Photon.* **18**, 1067–1075 (2024).
31. Rahman, M. S. S., Li, Y., Yang, X., Chen, S. & Ozcan, A. Massively parallel and universal approximation of nonlinear functions using diffractive processors. *eLight* **5**, 32 (2025).

32. Tian, Y. *et al.* Photonic transformer chip: interference is all you need. *PhotoniX* **6**, 45 (2025).

33. Kingma, D. P. & Ba, J. Adam: A Method for Stochastic Optimization. (2014) doi:10.48550/ARXIV.1412.6980.

34. Alpaydin, E. & Kaynak, C. Optical Recognition of Handwritten Digits. UCI Machine Learning Repository https://doi.org/https://doi.org/10.24432/C50P49 (1998).

35. Huang, C. *et al.* Demonstration of scalable microring weight bank control for large-scale photonic integrated circuits. *APL Photonics* **5**, 040803 (2020).

36. Feldmann, J. *et al.* Parallel convolutional processing using an integrated photonic tensor core. *Nature* **589**, 52–58 (2021).

37. Zhang, H. *et al.* An optical neural chip for implementing complex-valued neural network. *Nat Commun* **12**, 457 (2021).

38. Ashtiani, F., Geers, A. J. & Aflatouni, F. An on-chip photonic deep neural network for image classification. *Nature* **606**, 501–506 (2022).

39. Zhu, H. H. *et al.* Space-efficient optical computing with an integrated chip diffractive neural network. *Nat Commun* **13**, 1044 (2022).

40. Wu, T., Menarini, M., Gao, Z. & Feng, L. Lithography-free reconfigurable integrated photonic processor. *Nat. Photon.* **17**, 710–716 (2023).

41. Pai, S. *et al.* Experimentally realized in situ backpropagation for deep learning in photonic neural networks. *Science* **380**, 398–404 (2023).

42. Bandyopadhyay, S. *et al.* Single-chip photonic deep neural network with forward-only training. *Nat. Photon.* **18**, 1335–1343 (2024).

43. Delaney, M. *et al.* Nonvolatile programmable silicon photonics using an ultralow-loss $Sb_2Se_3$ phase

change material. *Sci. Adv.* **7**, eabg3500 (2021).

44. Ho, J., Jain, A. & Abbeel, P. Denoising diffusion probabilistic models. in *Proceedings of the 34th International Conference on Neural Information Processing Systems* (Curran Associates Inc., Red Hook, NY, USA, 2020).
45. Oord, A. van den *et al.* WaveNet: A Generative Model for Raw Audio. in *Speech Synthesis Workshop* (2016).
46. Ho, J. *et al.* Video Diffusion Models. in *Advances in Neural Information Processing Systems* (eds Koyejo, S. et al.) vol. 35 8633–8646 (Curran Associates, Inc., 2022).
47. Li, J., Tang, T., Zhao, W. X., Nie, J.-Y. & Wen, J.-R. Pre-Trained Language Models for Text Generation: A Survey. *ACM Comput. Surv.* **56**, 1–39 (2024).
48. Acharya, D. B., Kuppan, K. & Divya, B. Agentic AI: Autonomous Intelligence for Complex Goals—A Comprehensive Survey. *IEEE Access* **13**, 18912–18936 (2025).
49. Liang, Z., He, H., Yang, C. & Dai, B. Scaling Laws for Diffusion Transformers. in *The Fourteenth International Conference on Learning Representations* (2026).
50. Chen, S., Li, Y., Wang, Y., Chen, H. & Ozcan, A. Optical generative models. *Nature* **644**, 903–911 (2025).
51. Chen, Y. *et al.* All-optical synthesis chip for large-scale intelligent semantic vision generation. *Science* **390**, 1259–1265 (2025).
52. Ronneberger, O., Fischer, P. & Brox, T. U-Net: Convolutional Networks for Biomedical Image Segmentation. in *Medical Image Computing and Computer-Assisted Intervention – MICCAI 2015* (eds Navab, N., Hornegger, J., Wells, W. M. & Frangi, A. F.) 234–241 (Springer International Publishing, Cham, 2015).
53. LeCun, Y., Cortes, C. & Burges, C. J. C. The MNIST Database of Handwritten Digits.

https://yann.lecun.com/exdb/mnist/ (1998).

54. Salimans, T. *et al.* Improved Techniques for Training GANs. in *Advances in Neural Information Processing Systems* (eds Lee, D., Sugiyama, M., Luxburg, U., Guyon, I. & Garnett, R.) vol. 29 (Curran Associates, Inc., 2016).
55. Heusel, M., Ramsauer, H., Unterthiner, T., Nessler, B. & Hochreiter, S. GANs Trained by a Two Time-Scale Update Rule Converge to a Local Nash Equilibrium. in *Advances in Neural Information Processing Systems* (eds Guyon, I. et al.) vol. 30 (Curran Associates, Inc., 2017).
56. Sankaranarayanan, S., Balaji, Y., Jain, A., Lim, S. N. & Chellappa, R. Learning From Synthetic Data: Addressing Domain Shift for Semantic Segmentation. in *Proceedings of the IEEE Conference on Computer Vision and Pattern Recognition (CVPR)* (2018).
57. Lu, J., Benea-Chelmus, I.-C., Ginis, V., Ossiander, M. & Capasso, F. Cascaded-mode interferometers: Spectral shape and linewidth engineering. *Sci. Adv.* **11**, eadt4154 (2025).
58. Yanagimoto, R. *et al.* Programmable on-chip nonlinear photonics. *Nature* https://doi.org/10.1038/s41586-025-09620-9 (2025) doi:10.1038/s41586-025-09620-9.
59. Nagi, J. *et al.* Max-pooling convolutional neural networks for vision-based hand gesture recognition. in *2011 IEEE International Conference on Signal and Image Processing Applications (ICSIPA)* 342–347 (IEEE, Kuala Lumpur, Malaysia, 2011). doi:10.1109/ICSIPA.2011.6144164.

## Acknowledgments

This work was supported by the National Research and Development Program of China (2023YFB2804702); National Natural Science Foundation of China (NSFC) (62550072, 62341508, 625B2113, 62405185, and 62505175); Shanghai Science and Technology Innovation Action Plan (25LN3201000, 25JD1405500 and 24JD1401500); Shanghai Municipal Science and Technology Major Project. We also thank the Center for Advanced Electronic Materials and Devices (AEMD) of Shanghai Jiao Tong University (SJTU) for fabrication support.

## Author contributions

X.H.G initiated the project. Z.Y.Z performed the calculation and simulation. Z.Y.Z and X.H.G. designed the experiments. Z.Y.Z and X.H fabricated samples. Z.Y.Z, Z.J.Q and Y.Z carried out the measurements. Z.Y.Z, Z.J.Q, X.H, Z.Y, J.L.X, Y.L.C, C.J.X, Y.C.Y, T.L, Y.K.S and X.H.G analyzed the results and wrote the manuscript. X.H.G supervised the project.

## Competing interests

The authors declare no competing interests.

**Correspondence and requests for materials** should be addressed to Xuhan Guo.